\documentclass[unpublished,a4paper,twocolumn,11pt]{quantumarticle}
\pdfoutput=1
\usepackage[utf8]{inputenc}
\usepackage{textgreek}
\usepackage[english]{babel}
\usepackage[T1]{fontenc}
\usepackage{amsmath}
\usepackage{hyperref}

\usepackage{tikz}
\usepackage{lipsum}
\usepackage[numbers,sort&compress]{natbib}

\makeatletter
\providecommand{\@afterenddocumenthook}{}
\makeatother

\begin{document}

\title{Coherence protection of a silicon hole spin qubit with phase-modulated microwave driving}

\author{Sayyid Irsyadul Ibad}
\affiliation{Department of Electrical and Electronic Engineering, Institute of Science Tokyo, 
Meguro, Tokyo 152-8552, Japan}
\email{sayyid.i.5937@m.isct.ac.jp}
\orcid{0009-0001-9051-320X}

\author{Yusuke Sato}
\email{yusukesato114@gmail.com}
\orcid{0009-0003-9217-8569}
\affiliation{Department of Electrical and Electronic Engineering, Institute of Science Tokyo, 
Meguro, Tokyo 152-8552, Japan}

\author{Takuma Kuno}
\affiliation{Department of Electrical and Electronic Engineering, Institute of Science Tokyo, 
Meguro, Tokyo 152-8552, Japan}
\affiliation{Research and Development Group, Hitachi, Ltd., Kokubunji, Tokyo 185-8601, Japan}

\author{Itaru Yanagi}
\affiliation{Research and Development Group, Hitachi, Ltd., Kokubunji, Tokyo 185-8601, Japan}

\author{Toshiyuki Mine}
\affiliation{Research and Development Group, Hitachi, Ltd., Kokubunji, Tokyo 185-8601, Japan}

\author{Ryuta Tsuchiya}
\affiliation{Research and Development Group, Hitachi, Ltd., Kokubunji, Tokyo 185-8601, Japan}

\author{Digh Hisamoto}
\affiliation{Research and Development Group, Hitachi, Ltd., Kokubunji, Tokyo 185-8601, Japan}

\author{Hiroyuki Mizuno}
\affiliation{Research and Development Group, Hitachi, Ltd., Kokubunji, Tokyo 185-8601, Japan}

\author{Raisei Mizokuchi}
\affiliation{Department of Electrical and Electronic Engineering, Institute of Science Tokyo, 
Meguro, Tokyo 152-8552, Japan}

\author{Jun Yoneda}
\affiliation{Academy of Super Smart Society, Institute of Science Tokyo, 
Meguro, Tokyo 152-8552, Japan}
\affiliation{Department of Advanced Materials Science, University of Tokyo, Kashiwa, 
Chiba 277-8561, Japan}

\author{Tetsuo Kodera}
\email{kodera.t.a173@m.isct.ac.jp}
\orcid{0000-0002-3740-3896}
\affiliation{Department of Electrical and Electronic Engineering, Institute of Science Tokyo, 
Meguro, Tokyo 152-8552, Japan}

\begin{abstract}
   Hole spins in silicon quantum dots are a promising platform for quantum computing due to their strong intrinsic spin-orbit coupling (SOC), which enables fast, all-electrical control. However, this coupling also increases their susceptibility to charge noise, thereby limiting coherence times. Moreover, holes in silicon are also affected by hyperfine interactions with residual nuclear spins in the silicon substrate, introducing a non-negligible source of low-frequency noise. Here, we implement a phase-modulated concatenated continuous driving (CCD) technique for hole spin qubits to suppress low-frequency noise through microwave phase modulation. This approach stabilizes Rabi oscillations and extends the oscillation decay time compared to the conventional method. Furthermore, by defining a qubit in the CCD frame, we achieve coherent control while simultaneously protecting the qubit from noise, confirming coherence protection during gate operations. These results establish a promising approach to coherence protection in semiconductor hole spin qubits.
 
\end{abstract}

Spin qubits in silicon quantum dots are promising building blocks for scalable quantum computing because they are compatible with existing complementary metal-oxide-semiconductor (CMOS) technology, facilitating large-scale integration~\cite{veldhorst2017silicon,li2018crossbar,gonzalez2021scaling,kunne2024spinbus,siegel2024towards}. Among them, hole spin qubits have gained significant interest due to their strong intrinsic spin--orbit coupling (SOC), which allows fast and fully electrical control without the need for additional structures~\cite{bulaev2007electric,crippa2018electrical,camenzind2022hole,fang2023recent,Vorreiter2026precision} such as microstrip lines~\cite{veldhorst2014addressable} or micromagnets~\cite{yoneda2018quantum}. However, SOC also enhances sensitivity to charge noise, which typically has a $1/f$-like spectrum that dominates at low frequencies~\cite{culcer2009dephasing,piot2022single,struck2020low}. Moreover, recent studies have also shown that, even in hole systems, hyperfine interactions with nuclear spins introduce a non-negligible source of low-frequency noise~\cite{piot2022single,hendrickx2024sweet}. Together, these noise sources constitute the primary limitation on hole spin coherence times.

Noise decoupling techniques are often used to enhance coherence times by suppressing low-frequency noise through the application of timed control pulses or continuous driving fields. Dynamical decoupling methods like Hahn echo~\cite{hahn1950spin} and Carr--Purcell--Meiboom--Gill (CPMG) protocols~\cite{carr1954effects,viola1998dynamical,vandersar2012decoherence,zhang2014protected} effectively suppress dephasing by adding refocusing pulses during idle operations. Another well-known decoupling approach involves continuously applying modulated microwave (MW) schemes (dressing qubits) such as the sinusoidally modulated, always rotating, and tailored (SMART)~\cite{hansen2021pulse,hansen2022implementation} technique and concatenated continuous driving (CCD) technique~\cite{ramsay2023coherence,wang2020coherence,kuno2026robust}. In contrast to pulse-based dynamical decoupling, continuous driving enables noise decoupling not only during idle periods but also during gate operations, protecting qubits against both dephasing noise and driving noise. While these approaches have successfully extended the coherence times of electron spin qubits, extending these continuous driving benefits to hole spin qubits remains an open experimental challenge.

\begin{figure*}[t]
  \centering
  \includegraphics[width=\textwidth]{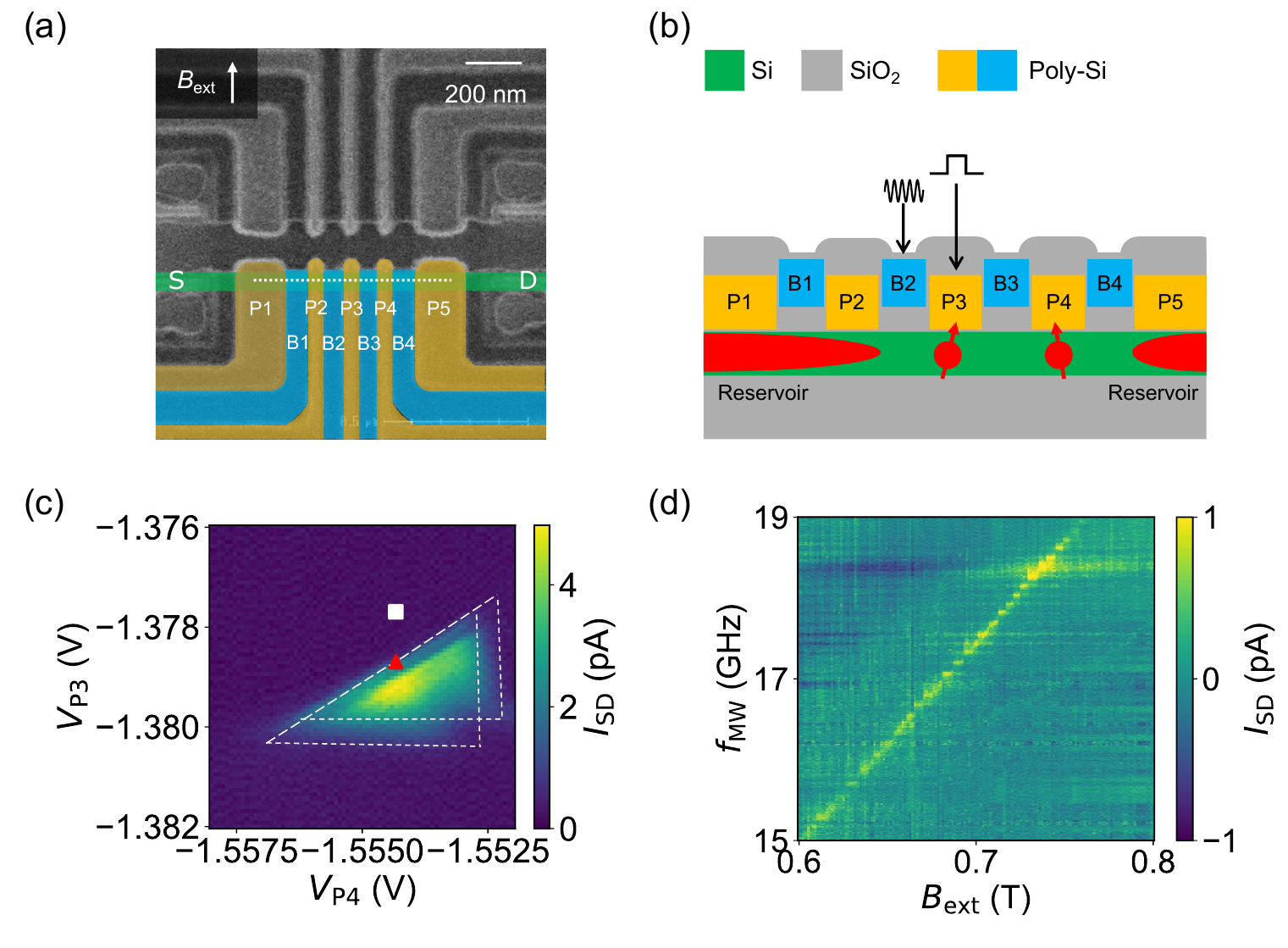}
  \caption{\textbf{Device setup, PSB and EDSR.} \textbf{(a)} False-colored scanning electron micrograph of the FDSOI quantum dot array consisting of a natural Si channel with p-type source (S) and drain (D) reservoirs, and two-layer poly-Si gates: plunger (P1--5) and barrier (B1--4) gates. An in-plane magnetic field $B_{\text{ext}}$ is applied perpendicularly to the DQD channel. \textbf{(b)} Cross-sectional schematic along the white dashed line in (a). The DQD is formed under P3 and P4, while two large two-dimensional hole gases serving as reservoirs are formed under P1, B1, and P2, and under P5, respectively. MW and pulse signals are applied to B2 and P3, respectively. \textbf{(c)} DQD system showing a typical bias triangle in the transport current $I_{\text{SD}}$. The red triangle and white square mark the PSB regime and Coulomb blockade regime, respectively. \textbf{(d)} DQD transport current $I_{\text{SD}}$ as a function of MW frequency $f_{\text{MW}}$ and magnetic field $B_{\text{ext}}$ in the PSB regime. A resonance line is observed when $f_{\text{MW}}$ matches the Larmor frequency. For clarity, the $f_{\text{MW}}$- and $B_{\text{ext}}$-dependent current offset is subtracted from the data.}
  \label{fig:figure1}
\end{figure*}

In this work, we implement the phase-modulated CCD protocol, a variant of the CCD protocol, for hole spins in silicon quantum dots and demonstrate its effectiveness. We employ the phase-modulated CCD protocol because it provides robustness against MW amplitude instability~\cite{ramsay2023coherence,kuno2026robust,cohen2017continuous} and mitigates the unwanted $g$-factor modulation inherent to hole systems~\cite{crippa2018electrical}, making it well suited for hole-spin systems. We demonstrate hole spin control via electric-dipole spin resonance (EDSR) and characterize the coherence times through Rabi, Ramsey, and Hahn echo measurements. We then apply the phase-modulated CCD protocol and demonstrate improved coherence through stabilized Rabi oscillations. We further define and coherently control a hole spin qubit in the CCD-dressed frame, demonstrating robust noise protection and coherence preservation during qubit operations.

The device used in this study is fabricated using fully depleted silicon-on-insulator (FDSOI) technology, which offers high compatibility with CMOS processes~\cite{lee202216}. The device consists of a natural Si channel with p-type source (S) and drain (D) reservoir terminals, and two poly-Si gate layers (plunger gate layer and barrier gate layer) are formed on top of the Si channel as depicted in Figures~\ref{fig:figure1}a and \ref{fig:figure1}b. We form a double quantum dot (DQD) under gates P3 and P4 and the transport current $I_{\text{SD}}$ is measured to characterize the DQD. We apply pulses and MW signals to P3 and B2 gates, respectively, to enable spin control. The device is cooled to $100\text{ mK}$ and an in-plane external magnetic field of $B_{\text{ext}} = 700\text{ mT}$ is applied perpendicularly to the DQD channel. Figure \ref{fig:figure1}c shows the bias triangles used in the experiments with a source-drain voltage of $1.5\text{ mV}$.

Spin-to-charge conversion is achieved via Pauli spin blockade (PSB)~\cite{ono2002current,lai2011pauli,seedhouse2021pauli}, where transport between two dots can be blocked by spin conservation. PSB can be lifted by rotating a hole spin in one of the two dots using EDSR with a MW field resonant with the Zeeman splitting. Figure \ref{fig:figure1}d shows the DQD current as a function of magnetic field $B_{\text{ext}}$ and MW frequency $f_{\text{MW}}$ in the PSB regime. The $f_{\text{MW}}$- and $B_{\text{ext}}$-dependent current offset is subtracted to enhance the visibility. A resonance line is observed, originating from the EDSR-induced hole spin rotation. From the slope of the resonance line, we obtain an effective hole $g$-factor of $1.79$.

\begin{figure*}[t]
  \centering
  \includegraphics[width=\textwidth]{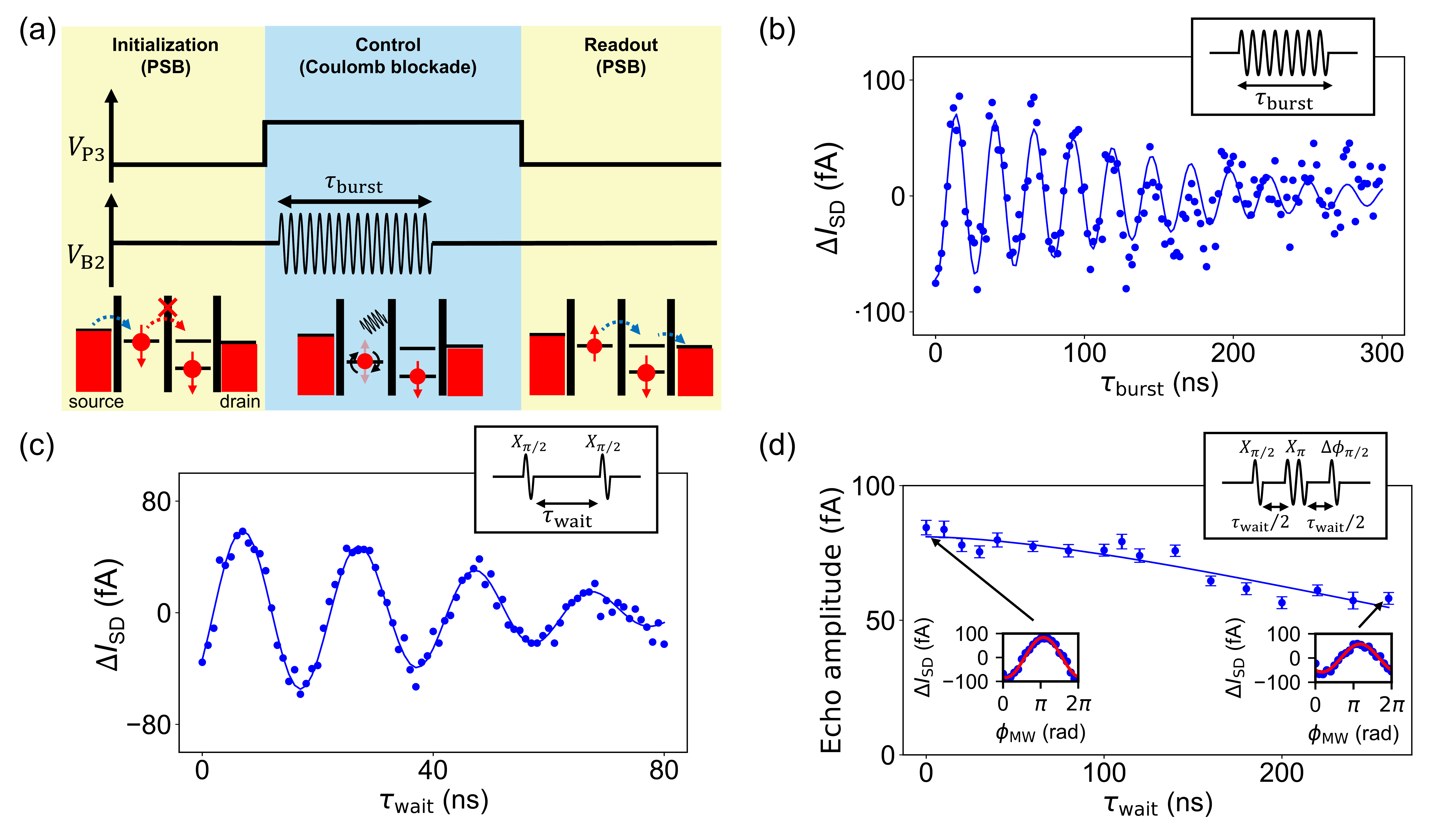}
  \caption{\textbf{Coherent control of a single-hole spin.} \textbf{(a)} Schematic of the pulse scheme for coherent single-spin control, showing initialization and readout in the PSB regime and manipulation in the Coulomb blockade regime. \textbf{(b)} Rabi oscillation and its pulse scheme (top inset). The data are fitted to a decaying sinusoidal function (solid line). The transport current data $\Delta I_{\text{SD}}$ has been corrected by subtracting the burst time-dependent offset. \textbf{(c)} Ramsey oscillation at a frequency detuning of $50\text{ MHz}$ and its pulse scheme (top inset). The data are fitted with a Gaussian decaying oscillation (solid line). $\Delta I_{\text{SD}}$ is integrated for $1\text{ s}$ at each of the 81 wait-time points. The same sequence is repeated five times with an averaging interval of about $50\text{ s}$, and the average of these five data is plotted. \textbf{(d)} Hahn echo amplitude versus wait time $\tau_{\text{wait}}$, with the pulse scheme shown in the top inset. As represented in the two lower panels, the amplitude for each $\tau_{\text{wait}}$ corresponds to the oscillation amplitude of the transport current observed when the phase of the final echo pulse is varied. The solid line is a fit to $A \exp\left( -\left(\tau_{\text{wait}} / T_2^{\text{H}}\right)^{\alpha+1} \right)$ with $\alpha = 0.6$, where the standard deviation error bars are derived from the fitting.}
  \label{fig:figure2}
\end{figure*}

Coherent spin control is achieved using a three-stage pulse sequence (Figure \ref{fig:figure2}a). First, spins are initialized via PSB. Next, a resonant MW burst is applied in the Coulomb blockade regime to rotate the spin. Finally, spin readout is performed by returning to the PSB regime. This cycle with a period of less than $1~\mu\text{s}$ is repeated a million times yielding measurable current $I_{\text{SD}}$. By varying the MW burst duration $\tau_{\text{burst}}$, we observe Rabi oscillations with a Rabi frequency of $f_{\text{R}} = 38\text{ MHz}$ and a decay time of $T_2^{\text{R}} = (177 \pm 20)\text{ ns}$ (Figure \ref{fig:figure2}b). This relatively short decay time is attributed to the environmental noise and driving noise.

To evaluate the dephasing effect induced by environmental noise, we perform Ramsey and Hahn echo experiments. In the Ramsey experiment, two $\pi/2$ pulses are applied, separated by a variable wait time $\tau_{\text{wait}}$, during which the qubit freely evolves at a frequency offset of $\Delta f_{\text{MW}} \approx 50\text{ MHz}$. Figure \ref{fig:figure2}c shows the resulting transport current oscillations. By fitting the results to $A \exp\left( -\left( \frac{\tau_{\text{wait}}}{T_2^*} \right)^2 \right) \cos(\Delta f_{\text{MW}} \tau_{\text{wait}} + \theta)$, we extract a Ramsey coherence time $T_2^*$ of $(58 \pm 2)\text{ ns}$. The coherence time can be extended by using Hahn echo sequence where a $\pi$ pulse is applied in between the $\pi/2$ pulses which refocuses slow phase drift induced by low-frequency noise. Figure \ref{fig:figure2}d displays the Hahn echo measurement results, where each data point represents the oscillation amplitude of the transport current, obtained by sweeping the rotation axis of the final $\pi/2$ pulse. By fitting the data to $A \exp\left( -\left( \frac{\tau_{\text{wait}}}{T_2^H} \right)^{\alpha+1} \right)$, we obtain $\alpha = 0.6$ and the Hahn-echo coherence time $T_2^H = (467 \pm 21)\text{ ns}$.

The measured $T_2^{\text{H}}$ is relatively short compared to the microsecond-scale coherence times typically observed in hole spin qubits~\cite{Vorreiter2026precision,piot2022single,hendrickx2024sweet}. This limitation is likely due to several factors, such as strong high-frequency charge noise or possibly due to the finite coupling of the quantum dots to the source and drain reservoirs, which is relatively large in the transport measurement experiments~\cite{hendrickx2020single}. The latter possibility is supported by the comparable values of our measured coherence times and those reported previously using similar transport measurement techniques~\cite{maurand2016cmos,watzinger2018germanium}.

We note that the enhancement of the coherence time by the echo sequence, $T_2^H / T_2^* \approx 8$, indicates that low-frequency noise is the dominant factor limiting $T_2^*$. The low-frequency noise can possibly originate from several mechanisms. One possible source is the residual hyperfine interaction with $^{29}\text{Si}$ nuclear spins, which is known to limit coherence even in hole systems~\cite{Vorreiter2026precision,piot2022single}. However, our estimation based solely on the hyperfine dephasing effect in natural silicon~\cite{bosco2021fully} yields an expected $T_2^*$ on the order of microseconds (see Supporting Information~\ref{supp:hyperfine_T2star}), indicating that other additional mechanisms likely contribute to the observed decoherence. A highly probable candidate is low-frequency charge noise, which typically exhibits a $1/f$-like spectrum and couples strongly to the spin via the strong SOC~\cite{piot2022single,malkoc2022charge}. Such charge noise may originate from charge traps, slow gate-voltage fluctuations, and, in transport-based measurements, slow electrostatic fluctuations associated with the source and drain reservoirs. Ultimately, more detailed characterization is required to definitively identify the dominant noise source~\cite{piot2022single,hendrickx2024sweet,yoneda2023noise}.

Given that the device exhibits short coherence times limited by low-frequency noise, we now investigate the ability of the phase-modulated CCD protocol to suppress low-frequency noise, thereby extending the coherence times. We realize the CCD waveform by modulating the phase of the MW as 
\begin{multline}
v(t) = V_{\text{AC}} \cos\Big[ 2\pi f_{\text{MW}} t + \phi_{\text{MW}} \\
- \frac{2\epsilon_{\text{mod}}}{f_{\text{mod}}} \sin\left( 2\pi f_{\text{mod}} t - \theta_{\text{mod}} \right) \Big]
\label{eq:vt}
\end{multline}
where $f_{\text{MW}}$ is the MW frequency, $\phi_{\text{MW}}$ is the MW phase offset, and $\epsilon_{\text{mod}}$, $f_{\text{mod}}$, and $\theta_{\text{mod}}$ are the modulation amplitude, frequency, and phase of the CCD protocol, respectively.

The MW excitation generates an effective oscillating magnetic field through SOC. This effective field can be decomposed into transverse and longitudinal components with respect to the spin quantization axis. Assuming that the SOC-induced effective driving amplitudes are linear in the applied MW electric field, such that higher-order nonlinear contributions are neglected, and choosing the $x$-axis along the transverse component, the laboratory-frame Hamiltonian is given by

\begin{multline}
H_{\text{lab}} = \frac{h f_L}{2} \sigma_z + h \left( \lambda_{\perp} \sigma_x + \lambda_{\parallel} \sigma_z \right) \\
\cos \Big[ 2\pi f_{\text{MW}} t + \phi_{\text{MW}} \\
- \frac{2\epsilon_{\text{mod}}}{f_{\text{mod}}} \sin\left( 2\pi f_{\text{mod}} t - \theta_{\text{mod}} \right) \Big]
\label{eq:h_lab}
\end{multline}
where $\sigma_i$ ($i \in \{x,y,z\}$) are the Pauli spin matrices, $f_{\text{L}}$ is the qubit Larmor frequency, and $\lambda_{\perp}$ and $\lambda_{\parallel}$ are the amplitudes of the transverse and longitudinal fields, respectively, which depend on the MW amplitude $V_{\text{AC}}$ and the strength of the SOC. The transverse component leads to Rabi driving with a Rabi frequency of $f_{\text{R}} = \lambda_{\perp}$, whereas the longitudinal component periodically modulates the Larmor frequency without directly inducing spin rotation.

In the first rotating frame defined by the transformation $U(t) = \exp\left( -i \Phi \sigma_z / 2 \right)$, where $\Phi = 2\pi f_{\text{MW}} t - \frac{2\epsilon_{\text{mod}}}{f_{\text{mod}}} \sin\left( 2\pi f_{\text{mod}} t - \theta_{\text{mod}} \right)$, the longitudinal driving term oscillates at approximately $f_{\text{MW}}$ and therefore averages to zero under the rotating-wave approximation (RWA). The effective Hamiltonian is thus reduced to:
\begin{multline}
H_{\text{rot}}' = \frac{h\delta_L}{2} \sigma_z + \frac{hf_{\text{R}}}{2} \sigma_{\phi_{\text{MW}}} \\
+ h\epsilon_{\text{mod}} \cos\left( 2\pi f_{\text{mod}} t - \theta_{\text{mod}} \right) \sigma_z
\label{eq:h_rot1}
\end{multline}
where $\delta_{\text{L}} = f_{\text{L}} - f_{\text{MW}}$ is the frequency detuning and $\sigma_{\phi_{\text{MW}}} = \cos(\phi_{\text{MW}})\sigma_x + \sin(\phi_{\text{MW}})\sigma_y$. Under the above linear-response approximation and the RWA, the SOC enters only through the effective transverse and longitudinal driving amplitudes, $\lambda_{\perp}$ and $\lambda_{\parallel}$, without modifying the form of the rotating-frame Hamiltonian. Consequently, the effective Hamiltonian is formally identical to the conventional electron-spin CCD protocols~\cite{ramsay2023coherence,kuno2026robust}, demonstrating that the CCD mechanism remains fully applicable to hole-spin systems despite the presence of strong SOC.

In this first rotating frame, the Rabi drive term (second term) is a static field in the $x$-$y$-plane perpendicular to the frequency detuning term (first term) along the $z$-axis. This orthogonality provides protection against slow fluctuations in the frequency detuning. The CCD modulation term introduces a sinusoidal perturbation that modulates the Rabi field. This modulation provides refocusing of slow fluctuations in the Rabi driving, thereby offering an additional layer of protection against driving noise. This protection is particularly beneficial for systems with strong SOC since they are expected to exhibit stronger Rabi driving fluctuations due to electrical noise~\cite{liu2023ultrafast}.

\begin{figure*}[t]
  \centering
  \includegraphics[width=\textwidth]{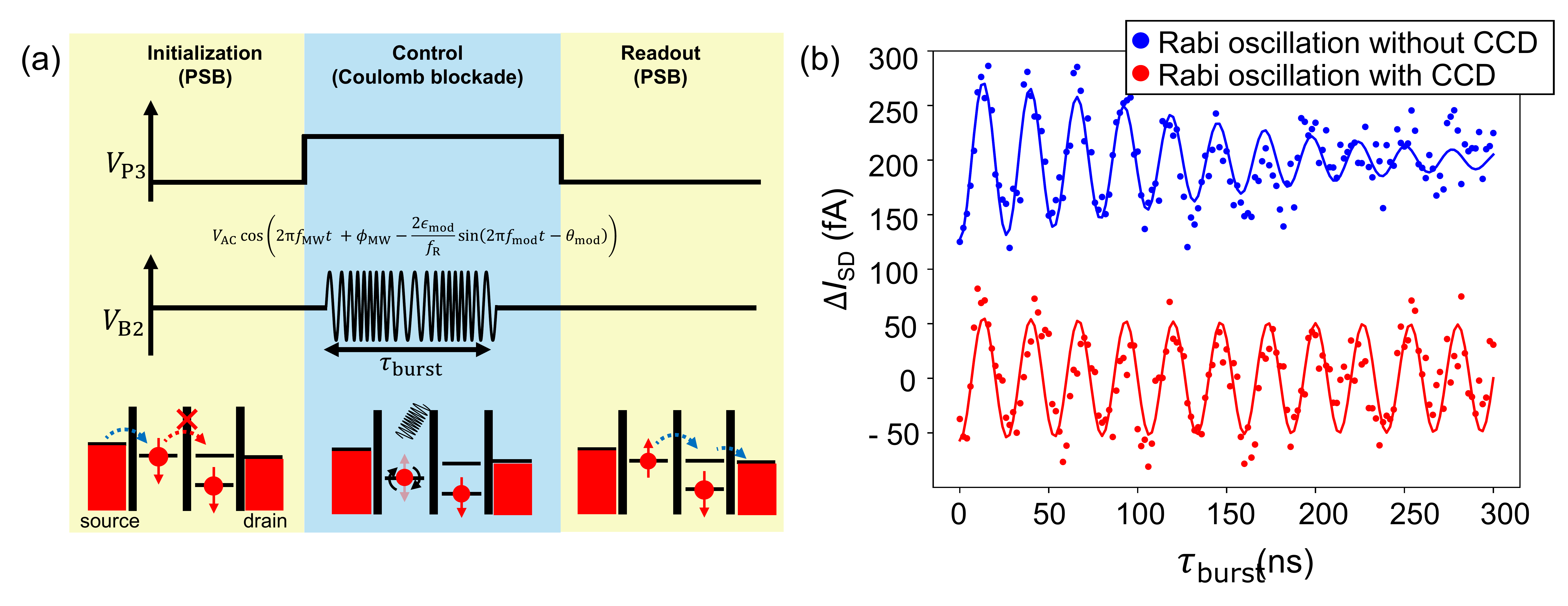}
  \caption{\textbf{Rabi experiments using the phase-modulated CCD technique.} \textbf{(a)} Schematic of the pulse scheme in which the MW burst is phase-modulated. \textbf{(b)} Comparison of Rabi oscillations without CCD (blue) and with CCD (red) fitted with a decaying sinusoidal function. The Rabi oscillation without CCD is offset by $200\text{ fA}$ for clarity.}
  \label{fig:figure3}
\end{figure*}

Figure \ref{fig:figure3} shows the pulse sequence of the Rabi experiments using the phase-modulated CCD, along with the corresponding results compared with the case without CCD. The experiments are performed with $\epsilon_{\text{mod}} = 0.2 f_{\text{R}}$, $\phi_{\text{MW}} = 0$, $\theta_{\text{mod}} = 0$, and $f_{\text{mod}} = f_{\text{R}} = 38\text{ MHz}$. The Rabi oscillations are stabilized at a frequency of $f_{\text{mod}}$, as shown in Figure \ref{fig:figure3}b, demonstrating the effectiveness of phase-modulated CCD in protecting the hole spin from slow fluctuations in frequency detuning and Rabi driving. We note that no decay is observed within the $300\text{ ns}$ measurement window, which is limited by the signal-to-noise ratio (SNR) of the transport current measurement (see Supporting Information~\ref{supp:transport_limitations}). This indicates that the Rabi coherence time $T_2^{\text{R-CCD}} \geq 300\text{ ns}$, representing at least a twofold improvement of coherence time compared to the conventional Rabi oscillations without CCD.

We then define a dressed qubit in the second rotating frame (CCD frame) that allows coherent control within the CCD-protected subspace, thereby enabling qubit operations while suppressing noise~\cite{ramsay2023coherence,kuno2026robust}. The Hamiltonian in the CCD frame is defined by further transforming the Hamiltonian in Eq.~\eqref{eq:h_rot1} using the transformation $U(t) = \exp\left( -\frac{i}{\hbar} \frac{h f_{\text{mod}}}{2} \sigma_{\phi_{\text{MW}}} t \right)$. Under the resonance condition $\delta_{\text{L}} = 0$, using the RWA, the Hamiltonian in the second rotating frame becomes:
\begin{multline}
H_{\text{rot}}'' = \frac{h\delta_{\text{mod}}}{2} \sigma_{\phi_{\text{MW}}} \\
+ \frac{h\epsilon_{\text{mod}}}{2} \left[ \cos(\theta_{\text{mod}}) \sigma_z + \sin(\theta_{\text{mod}}) \sigma_{\phi_{\text{MW}} + \pi/2} \right]
\label{eq:h_rot2}
\end{multline}
where the first term corresponds to the detuning between the Rabi frequency and the modulation frequency $\delta_{\text{mod}} = f_{\text{R}} - f_{\text{mod}}$, while the second term represents the driving term originating from the phase modulation.

At $\delta_{\text{mod}} = 0$, $\phi_{\text{MW}} = 0$, and $\theta_{\text{mod}} = 0$, the Hamiltonian reduces to $H_{\text{rot}}'' = \frac{h\epsilon_{\text{mod}}}{2}\sigma_z$. In this case, the effective field points along the $z$-axis, leading to precession around the $z$-axis at a frequency of $\epsilon_{\text{mod}}$. This corresponds to the idle mode of the dressed qubit. The eigenstates in this case are $|0''\rangle = \cos\left(\frac{2\pi f_{\text{mod}} t}{2}\right)|0\rangle - i \sin\left(\frac{2\pi f_{\text{mod}} t}{2}\right)|1\rangle$ and $|1''\rangle = i \sin\left(\frac{2\pi f_{\text{mod}} t}{2}\right)|0\rangle - \cos\left(\frac{2\pi f_{\text{mod}} t}{2}\right)|1\rangle$, with $|0\rangle$ and $|1\rangle$ being the bare qubit states corresponding to spin down $|\!\downarrow\rangle$ and spin up $|\!\uparrow\rangle$, respectively. When mapped back to the lab frame, these dressed eigenstates manifest as bare qubit Rabi oscillations. Importantly, the $|0''\rangle$ and $|1''\rangle$ states oscillate completely out of phase relative to each other, allowing us to read out and distinguish the two dressed eigenstates by measuring the phase of the resulting Rabi oscillations~\cite{ramsay2023coherence}.

Furthermore, by fixing the total MW burst time to integer multiples of the half modulation period $t_{\text{burst}} = n / (2 f_{\text{mod}})$, we can achieve matched readout conditions, where the dressed states align directly with the bare qubit states~\cite{kuno2026robust}. For even integers $n$, the dressed states map directly as $|0''\rangle = |0\rangle$ and $|1''\rangle = |1\rangle$. Conversely, for odd integers $n$, the CCD dressed states are inversely matched with the bare qubit states such that $|0''\rangle = |1\rangle$ and $|1''\rangle = |0\rangle$. By utilizing this mapping in combination with the PSB method, the dressed qubit states can be directly read out via changes in the transport current.

\begin{figure*}[t]
  \centering
  \includegraphics[width=\textwidth]{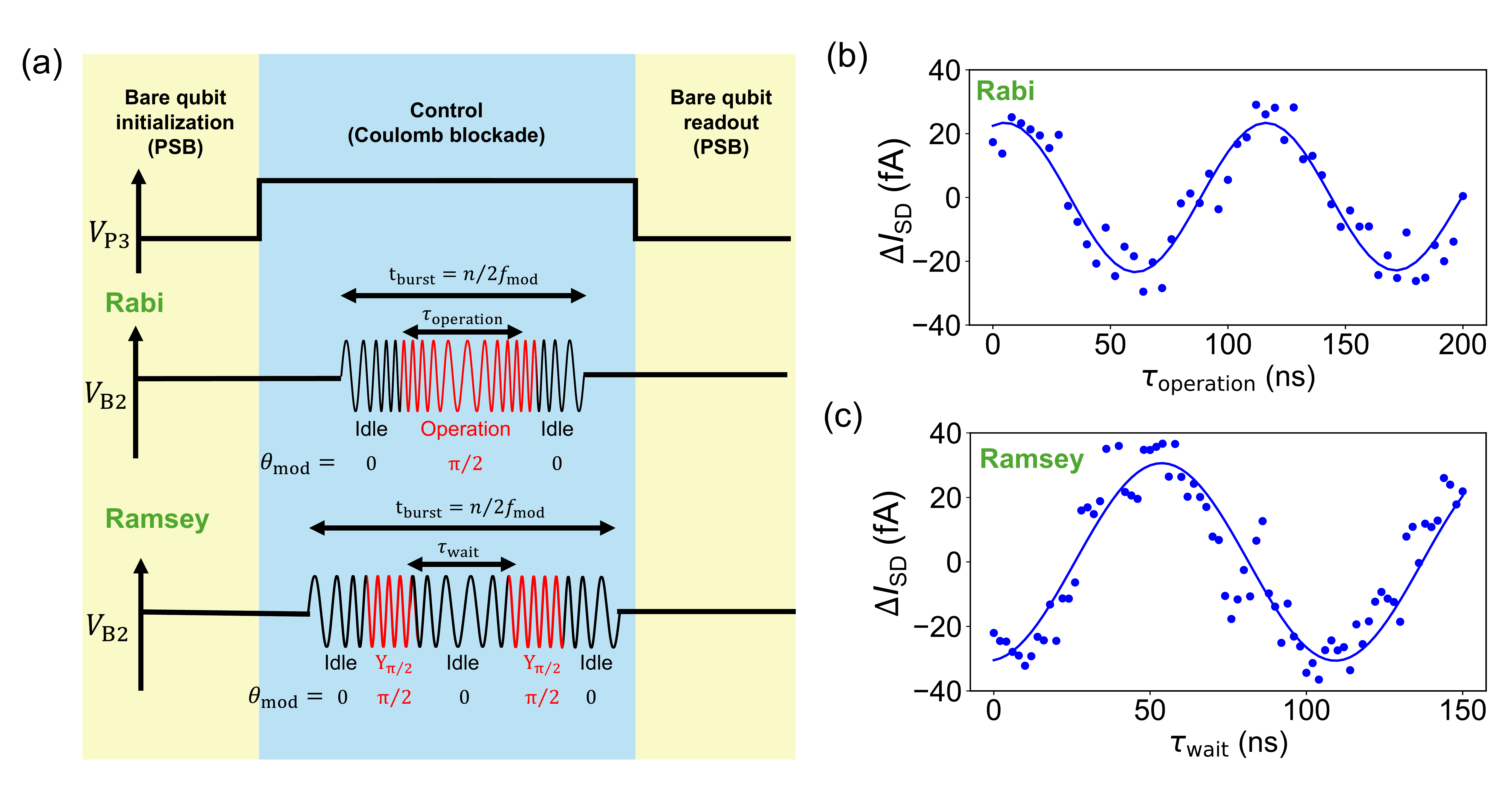}
  \caption{\textbf{Control of CCD-dressed qubits.} \textbf{(a)} Pulse schemes for Rabi, Ramsey, and two-axis control experiments. The idle and operation modes are controlled by setting the modulation phase to $\theta_{\text{mod}} = 0$ and $\theta_{\text{mod}} = \pi / 2$, respectively. \textbf{(b)} Rabi oscillations of a CCD-dressed qubit, fitted with a sinusoidal function. \textbf{(c)} Ramsey oscillations of the CCD-dressed qubit, fitted with a sinusoidal function. All data are obtained with the MW burst time fixed at $t_{\text{burst}} = n / (2 f_{\text{mod}})$ to satisfy the readout-matching condition, with $n = 23$ and $f_{\text{mod}} = 38\text{ MHz}$.}
  \label{fig:figure4}
\end{figure*}

Coherent control of the dressed qubit can be achieved by setting $\theta_{\text{mod}} = \pi / 2$~\cite{ramsay2023coherence,kuno2026robust}. Under this condition, the Hamiltonian in the CCD frame reduces to $H_{\text{rot}}'' = \frac{h\epsilon_{\text{mod}}}{2} \sigma_y$, which corresponds to rotations about the $y$-axis at a frequency of $\epsilon_{\text{mod}}$. To demonstrate coherent control of the dressed qubit, we perform Rabi and Ramsey experiments of the dressed qubit using the pulse sequence shown in Figure \ref{fig:figure4}a. For the Rabi experiment, the sequence begins in the PSB configuration to initialize the bare qubit, followed by pulsing to the Coulomb blockade regime. The bare qubit state is then mapped onto the dressed qubit state by applying an idle mode with an idle time of two bare qubit Rabi cycles ($t_{\text{idle}} = 2 / f_{\text{R}}$), ensuring that the frequency of the dressed system is well defined~\cite{ramsay2023coherence}. We then apply a $y$-axis rotation for a variable operation time $\tau_{\text{operation}}$, followed by another idle mode with a duration adjusted such that the total MW burst time satisfies a readout matching condition, $t_{\text{burst}} = n / (2 f_{\text{mod}})$. Finally, the system is pulsed back to the PSB region for qubit readout. For the Ramsey sequence, the continuous $y$-axis rotation is replaced by two $\pi / 2$ pulses separated by an idle mode with a variable waiting time $\tau_{\text{wait}}$. The first $\pi / 2$ pulse initializes the dressed qubit into a superposition state. During the subsequent idle period, the qubit undergoes coherent rotation around the $z$-axis, and the second $\pi / 2$ pulse projects the qubit onto the measurement basis for readout.

Figures \ref{fig:figure4}b and \ref{fig:figure4}c show the Rabi oscillation and Ramsey oscillation of the dressed qubit, respectively, with modulation amplitude $\epsilon_{\text{mod}} = 0.2 f_{\text{R}}$ and frequency $f_{\text{mod}} = f_{\text{R}} = 38\text{ MHz}$. The observed Rabi and Ramsey oscillations correspond to coherent rotations around the dressed $y$-axis and dressed $z$-axis, respectively. Notably, neither the Rabi nor the Ramsey oscillations exhibit observable decay within the experimental time window, indicating that the coherence times of the dressed qubit exceed the measurement duration ($T_2^{\text{DQ-Rabi}} > 200\text{ ns}$, $T_2^{\text{DQ-Ramsey}} > 150\text{ ns}$). The absence of observable decay demonstrates that phase-modulated CCD effectively protects the qubit against decoherence induced by low-frequency noise over the measured timescales. The noise protection provided by phase-modulated CCD can be understood in terms of its filter function, which shifts the qubit sensitivity to detuning noise and Rabi driving noise from the low-frequency regime to the higher-frequency regime, where the noise spectral density is typically lower. The shift of the noise sensitivity is determined by the modulation parameters $\epsilon_{\text{mod}}$ and $f_{\text{mod}}$ (see Supporting Information~\ref{supp:filter_function}).

We notice that both dressed qubit Rabi and Ramsey frequencies extracted from the experiments are $8.8\text{ MHz}$. These values deviate slightly from the expected value of $\epsilon_{\text{mod}} = 0.2 f_{\text{R}} = 7.6\text{ MHz}$. To investigate the origin of this discrepancy, we numerically time-evolved the qubit state under the full time-dependent laboratory-frame Hamiltonian. The experimentally obtained coherent oscillations are quantitatively reproduced by introducing an effective qubit-drive detuning $\delta_{\text{L}}$ of $-20\text{ MHz}$ (see Supporting Information~\ref{supp:numerical_simulation}). Several mechanisms could contribute to such effective detuning, including slow variations of the local electrostatic environment due to charge fluctuations or charge trapping, MW-induced electrostatic drift or heating, and MW-driven rectification of the quantum-dot position resulting from the anharmonicity of the confinement potential~\cite{takeda2018optimized}.

In conclusion, we have demonstrated the implementation of the phase-modulated CCD protocol to stabilize Rabi oscillations of a silicon hole spin qubit and achieve coherent control of a dressed qubit defined in the CCD frame. By applying phase-modulated CCD, we suppress the decay of Rabi oscillations in the first rotating frame and extend the coherence time, demonstrating robustness against fluctuations in both Rabi driving and detuning noise. Furthermore, by defining the qubit in the CCD frame, coherent control with improved coherence times can be performed while simultaneously protecting the qubit from low-frequency noise. These results represent the first demonstration of a CCD-dressed hole spin qubit and show that the CCD protocol effectively suppresses the effects of low-frequency noise in hole spin qubits. 

Looking ahead, the improvement in coherence time can be further quantified in terms of qubit gate fidelity via randomized benchmarking~\cite{muhonen2015quantifying}, which can be enabled through higher-SNR readout techniques such as charge sensing and RF reflectometry~\cite{schoelkopf1998radio,noiri2020radio}. Furthermore, combining CCD with sweet-spot operation~\cite{piot2022single,hendrickx2024sweet} may provide a promising direction for further improving the coherence and fidelity of hole spin qubits. Sweet-spot operation can suppress the qubit susceptibility to charge noise at the device level, whereas CCD can dynamically mitigate remaining low-frequency fluctuations. Together, these approaches may provide a robust pathway toward achieving longer coherence times and higher-fidelity quantum control while preserving the advantages of strong SOC in hole spin qubits.

\section*{Author Information}

\subsection*{Corresponding Authors}
\begin{itemize}
    \item \textbf{Sayyid Irsyadul Ibad}$^\dagger$ -- Department of Electrical and Electronic Engineering, Institute of Science Tokyo, Meguro, Tokyo 152-8552, Japan; Email: \href{mailto:sayyid.i.5937@m.isct.ac.jp}{sayyid.i.5937@m.isct.ac.jp}
    \item \textbf{Yusuke Sato}$^\dagger$,$^\ddagger$ -- Department of Electrical and Electronic Engineering, Institute of Science Tokyo, Meguro, Tokyo 152-8552, Japan; Email: \href{mailto:yusukesato114@gmail.com}{yusukesato114@gmail.com}
    \item \textbf{Tetsuo Kodera} -- Department of Electrical and Electronic Engineering, Institute of Science Tokyo, Meguro, Tokyo 152-8552, Japan; Email: \href{mailto:kodera.t.a173@m.isct.ac.jp}{kodera.t.a173@m.isct.ac.jp}
\end{itemize}

\subsection*{Present Addresses}
$^\ddagger$Tokyo Electron Technology Solutions Ltd., Nirasaki, Yamanashi 407-0192, Japan

\subsection*{Author Contributions}
$^\dagger$S.I.I. and Y.S. contributed equally to this work.

The manuscript was written through contributions of all authors. S.I.I. and Y.S. performed the experiment and analyzed the data. S.I.I., Y.S., T.Ku., R.M., J.Y., and T.Ko. discussed the results. S.I.I., Y.S., R.M., J.Y., and T.Ko. contributed to the measurement setup. I.Y., T.M., D.H., and R.T. designed and fabricated the device. S.I.I. and Y.S. wrote the manuscript with input from all co-authors, and all authors have given approval to the final version of the manuscript. H.M. and T.Ko. supervised the project.

\section*{Acknowledgments}
This work was financially supported by JST Moonshot R\&D Grant Number JPMJMS256H, MEXT Quantum Leap Flagship Program (MEXT Q-LEAP) Grant Number JPMXS0118069228, JST CREST Grant Number JPMJCR24A1, JST FOREST Grant Number JPMJFR244D, JST ASPIRE Grant Number JPMJAP25A2, and JSPS KAKENHI Grants-in-Aid for Scientific Research Grant Numbers JP25K24574, JP26K01331, and JP23K17327.

\bibliographystyle{unsrtnat}
\bibliography{references}

\onecolumn\newpage
\appendix

\begin{center}
  \Huge \textbf{Supporting Information}
\end{center}
\vspace{1em}

\renewcommand{\thesection}{S\arabic{section}}
\setcounter{section}{0}

\renewcommand{\theequation}{SE\arabic{equation}}
\setcounter{equation}{0}

\renewcommand{\thefigure}{S\arabic{figure}}
\setcounter{figure}{0}
\renewcommand{\thetable}{S\arabic{table}}
\setcounter{table}{0}

\section{Estimation of $T_2^*$ limited by hyperfine interaction for holes in natural silicon}
\label{supp:hyperfine_T2star}
When the nuclear spins are left in their natural, unpolarized states, the Overhauser field follows a Gaussian distribution, and the characteristic time for qubit decay due to hyperfine interactions is given by~\cite{bosco2021fully}
\begin{equation}
\bar{\tau} = \frac{\hbar}{|A_k|} \sqrt{\frac{3N}{\nu I(I+1)}},
\label{eq:tau_bar}
\end{equation}
where $A_k$ denotes the hyperfine coupling parameter, $N$ is the number of atoms in the quantum dot, and $\nu$ is the abundance of the isotope having nuclear spin $I$. For holes in silicon with the $^{29}\text{Si}$ isotope, $A_k = 1.67\ \mu\text{eV}$, $\nu = 4.7\%$, and $I = 1/2$ are employed. To estimate $N$, we calculate the volume of the quantum dot, $V_{\text{dot}}$. Assuming a cylindrical geometry with a diameter of $l = 20.4\text{--}42.4\text{ nm}$ (which is determined by the singlet--triplet splitting) and a height~\cite{liles2021electrical} of $5\text{ nm}$, $V_{\text{dot}}$ is determined to be $\sim 1.6 \times 10^3\text{--}7 \times 10^3\text{ nm}^3$. Given the atomic density of silicon ($\sim 50\text{ atoms/nm}^3$), we find $N \sim 8 \times 10^4\text{--}4 \times 10^5$. Consequently, $\bar{\tau}$ is estimated to be $\sim 1\text{--}2\ \mu\text{s}$.

Due to heavy-hole--light-hole mixing in hole spins, the spin dynamics depend heavily on the directions of confinement and the external magnetic field. Consequently, the observed dephasing time is expressed as $T_2^* = \bar{\tau} / \sqrt{\sigma}$, where $\sigma$ is a dimensionless factor representing the electrically tunable anisotropy of the hyperfine interaction strength. As a result, $T_2^*$ is highly sensitive to both the confinement potential and the magnetic-field orientation. In the absence of anisotropy ($\sigma \approx 1$), $T_2^*$ is comparable to $\bar{\tau}$, yielding $T_2^* \approx 1\text{--}2\ \mu\text{s}$. However, depending on the degree of anisotropy, $T_2^*$ may differ substantially from $\bar{\tau}$. In particular, $T_2^*$ can be significantly enhanced at specific operating points (sweet spots)~\cite{hendrickx2024sweet}. Nevertheless, $\bar{\tau}$ provides the characteristic timescale set by hyperfine interactions, indicating that the hyperfine-limited $T_2^*$ is expected to be on the order of microseconds.

\section{Limitation of transport measurements}
\label{supp:transport_limitations}

\begin{figure*}[t]
  \centering
  \includegraphics[width=0.6\textwidth]{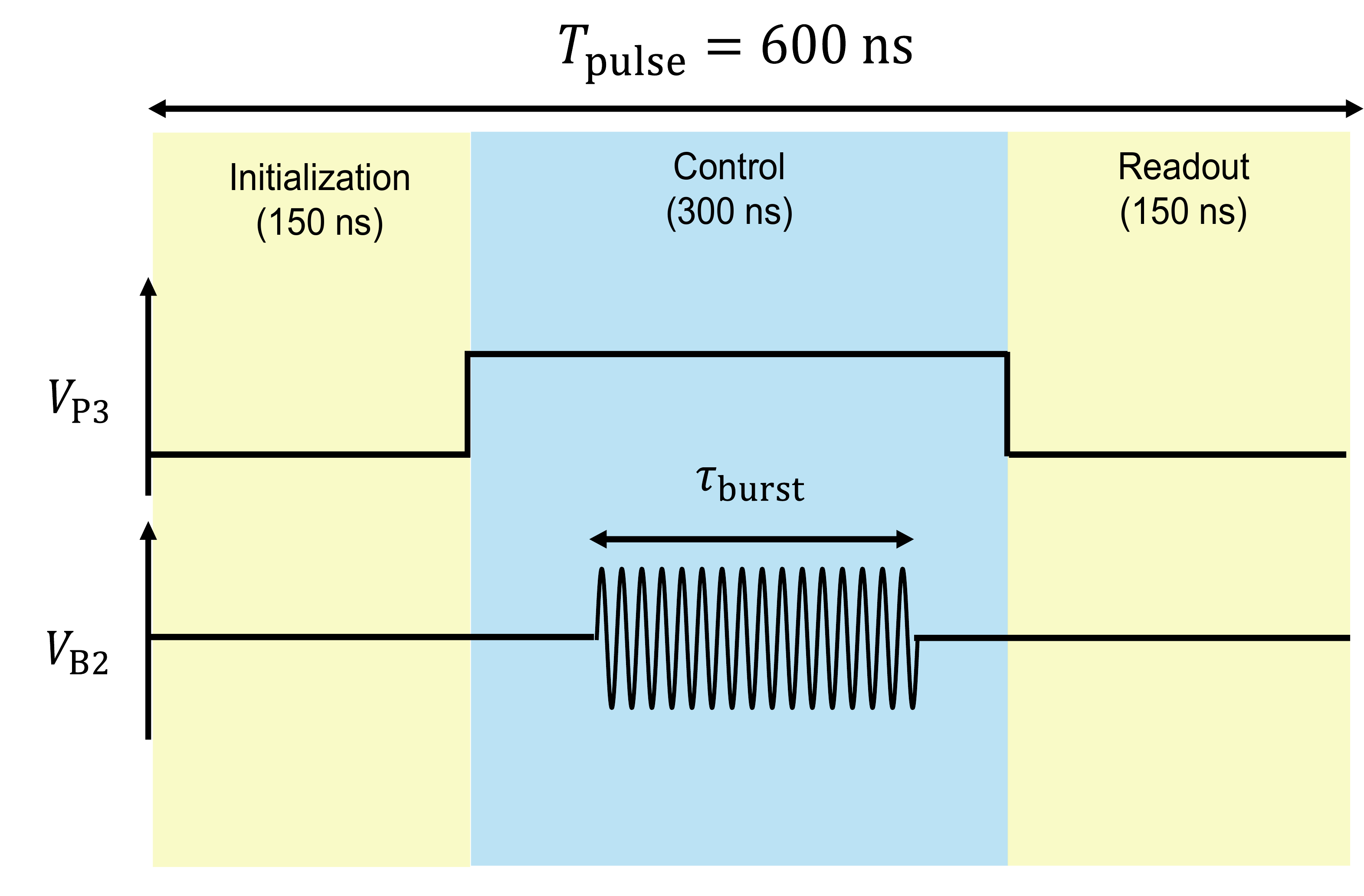}
  \caption{\textbf{Pulse schematic for spin operations used in the Rabi experiment.}}
  \label{fig:supp_pulse_scheme}
\end{figure*}

In this work, we measured changes in the spin states by applying a repeating two-level pulse a million times and averaging the total current over the pulse period, as shown in Figure~\ref{fig:supp_pulse_scheme}. The expected current change when a spin is fully flipped during EDSR measurement can be estimated as
\begin{equation}
\Delta I_{\text{DQD}} \approx \frac{e}{T_{\text{pulse}}},
\label{eq:delta_I_DQD}
\end{equation}
where $T_{\text{pulse}}$ corresponds to the period of the control pulse. In our experiments, we typically use $T_{\text{pulse}} = 600\text{ ns}$, which yields $\Delta I_{\text{DQD}} \approx 260\text{ fA}$. Using a longer $T_{\text{pulse}}$ will reduce $\Delta I_{\text{DQD}}$, making it difficult to measure due to the noise level in our measurement setup.

\section{Filter function derivation of phase-modulated CCD}
\label{supp:filter_function}

We derive the CCD filter function by defining the system and noise Hamiltonian in the lab frame~\cite{wang2020coherence}. For simplicity, we assume a microwave phase $\phi_{\text{MW}} = 0$ and set $h = 1$. Since the longitudinal field in Eq~\eqref{eq:h_lab} of the main text does not contribute to the spin rotation, we can neglect it, and the Hamiltonian becomes:
\begin{equation}
H_{\text{lab}} = \frac{f_{\text{L}} + \xi_{\delta}(t)}{2} \sigma_z + [f_{\text{R}} + \xi_{\text{R}}(t)] \cos\left( f_{\text{MW}} t - \frac{2(\epsilon_{\text{mod}} + \xi_{\text{mod}}(t))}{f_{\text{mod}}} \sin(2\pi f_{\text{mod}} t - \theta_{\text{mod}}) \right) \sigma_x,
\label{eq:H_lab}
\end{equation}
where $f_{\text{L}}$ is the qubit Larmor frequency, $f_{\text{R}}$ is the Rabi frequency, $f_{\text{MW}}$ is the microwave frequency, $f_{\text{mod}}$ is the modulation frequency, and $\theta_{\text{mod}}$ is the modulation phase. The operators $\sigma_i$ ($i \in \{x,y,z\}$) are the Pauli spin matrices, and $\xi_{\delta}(t)$, $\xi_{\text{R}}(t)$, and $\xi_{\text{mod}}(t)$ represent fluctuations in the detuning, the Rabi driving amplitude, and the modulation amplitude, respectively.

Transforming to the first rotating frame with $U(t) = \exp(-i \Phi \sigma_z / 2)$, where $\Phi = 2\pi f_{\text{MW}} t - (2\epsilon_{\text{mod}} / f_{\text{mod}}) \sin(2\pi f_{\text{mod}} t - \theta_{\text{mod}})$, and applying the rotating-wave approximation (RWA) under the resonance condition $f_{\text{R}} \ll f_{\text{MW}} = f_{\text{L}}$, we obtain:
\begin{equation}
H_{\text{rot}}' = \frac{f_{\text{R}}}{2} \sigma_x + \frac{\xi_{\text{R}}(t)}{2} \sigma_x + [\epsilon_{\text{mod}} + \xi_{\text{mod}}(t)] \cos(2\pi f_{\text{mod}} t - \theta_{\text{mod}}) \sigma_z + \frac{\xi_{\delta}(t)}{2} \sigma_z.
\label{eq:H_rot1}
\end{equation}

Transforming further to the second rotating frame with $U(t) = \exp(-i 2\pi f_{\text{mod}} t \sigma_x / 2)$, and assuming $\theta_{\text{mod}} = 0$, $f_{\text{mod}} = f_{\text{R}}$, and the RWA condition $\epsilon_{\text{mod}} < 2f_{\text{R}}$, we obtain:
\begin{equation}
H_{\text{rot}}'' = \frac{\epsilon_{\text{mod}}}{2} \sigma_z + \frac{\xi_{\delta}(t)}{2} [\sin(2\pi f_{\text{mod}} t) \sigma_y + \cos(2\pi f_{\text{mod}} t) \sigma_z] + \frac{\xi_{\text{R}}(t)}{2} \sigma_x + \frac{\xi_{\text{mod}}(t)}{2} \sigma_z.
\label{eq:H_rot2}
\end{equation}

We then move to the noise interaction picture~\cite{hansen2023accessing} with $U(t) = \exp(-i \epsilon_{\text{mod}} t \sigma_z / 2)$, yielding:
\begin{equation}
H_{\text{noise}}^{\text{int}} = H_{\delta}^{\text{int}} + H_{\text{R}}^{\text{int}} + H_{\text{mod}}^{\text{int}},
\label{eq:H_noise_int}
\end{equation}
with
\begin{align}
H_{\delta}^{\text{int}} &= \frac{\xi_{\delta}(t)}{2} \left[ -\sin(2\pi f_{\text{mod}} t) \sin(2\pi \epsilon_{\text{mod}} t) \sigma_x + \sin(2\pi f_{\text{mod}} t) \cos(2\pi \epsilon_{\text{mod}} t) \sigma_y + \cos(2\pi f_{\text{mod}} t) \sigma_z \right], \\
H_{\text{R}}^{\text{int}} &= \frac{\xi_{\text{R}}(t)}{2} \left[ \cos(2\pi \epsilon_{\text{mod}} t) \sigma_x + \sin(2\pi \epsilon_{\text{mod}} t) \sigma_y \right], \\
H_{\text{mod}}^{\text{int}} &= \frac{\xi_{\text{mod}}(t)}{2} \sigma_z.
\end{align}

From these expressions, we determine $y_{\alpha,\beta}(t)$, the modulation function of noise $\alpha \in \{\xi_{\delta}, \xi_{\text{R}}, \xi_{\text{mod}}\}$ along axis $\beta \in \{x, y, z\}$ on the Bloch sphere, and obtain the filter functions $F_{\alpha,\beta}(f,t)$ through:
\begin{equation}
F_{\alpha,\beta}(f,t) = \left| \int_0^t d\tau\, y_{\alpha,\beta}(\tau) \exp(i2\pi f\tau) \right|^2.
\label{eq:filter_def}
\end{equation}

The filter functions for detuning noise are:
\begin{align}
F_{\xi_{\delta},x}(f,t) &= \left| \int_0^t d\tau \left(-\sin(2\pi f_{\text{mod}} \tau) \sin(2\pi \epsilon_{\text{mod}} \tau)\right) \exp(i2\pi f\tau) \right|^2, \label{eq:F_delta_x} \\
F_{\xi_{\delta},y}(f,t) &= \left| \int_0^t d\tau \left(\sin(2\pi f_{\text{mod}} \tau) \cos(2\pi \epsilon_{\text{mod}} \tau)\right) \exp(i2\pi f\tau) \right|^2, \label{eq:F_delta_y} \\
F_{\xi_{\delta},z}(f,t) &= \left| \int_0^t d\tau \cos(2\pi f_{\text{mod}} \tau) \exp(i2\pi f\tau) \right|^2. \label{eq:F_delta_z}
\end{align}

For Rabi driving noise:
\begin{align}
F_{\xi_{\text{R}},x}(f,t) &= \left| \int_0^t d\tau \cos(2\pi \epsilon_{\text{mod}} \tau) \exp(i2\pi f\tau) \right|^2, \label{eq:F_R_x} \\
F_{\xi_{\text{R}},y}(f,t) &= \left| \int_0^t d\tau \sin(2\pi \epsilon_{\text{mod}} \tau) \exp(i2\pi f\tau) \right|^2, \label{eq:F_R_y}
\end{align}
and for modulation amplitude noise:
\begin{equation}
F_{\xi_{\text{mod}},z}(f,t) = \left| \int_0^t d\tau \frac{1}{2} \exp(i2\pi f\tau) \right|^2.
\label{eq:F_mod_z}
\end{equation}

\begin{figure*}[t]
  \centering
  \includegraphics[width=\textwidth]{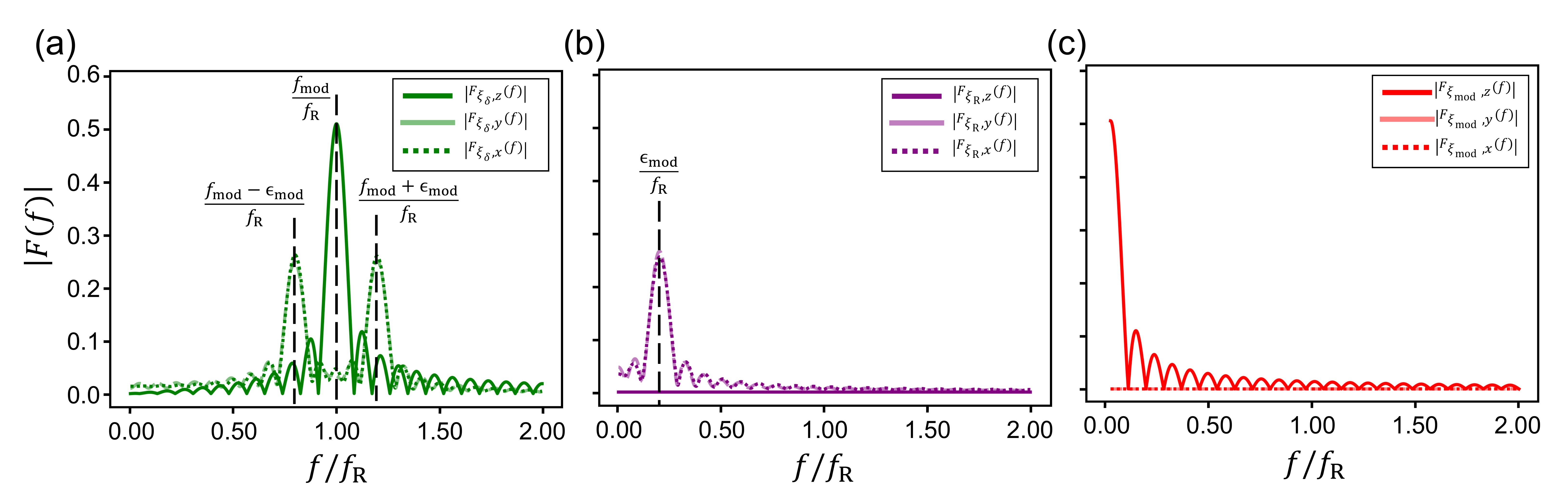}
  \caption{\textbf{Filter functions of CCD calculated for $f_{\text{mod}} = f_{\text{R}}$, $\epsilon_{\text{mod}} = 0.2 f_{\text{R}}$, and a total evolution time $t = 10/f_{\text{R}}$.} \textbf{(a)} Filter function for detuning noise, \textbf{(b)} filter function for driving noise, and \textbf{(c)} filter function for modulation noise. Solid, dashed, and dotted lines correspond to noise along the $z$-, $y$-, and $x$-axes, respectively. The peaks indicate the frequencies at which the CCD qubit is most sensitive to noise. The peak positions are determined by the CCD parameters, such as $\epsilon_{\text{mod}}$ and $f_{\text{mod}}$.}
  \label{fig:supp_filter_functions}
\end{figure*}

The filter functions for each noise component, with $\epsilon_{\text{mod}} = 0.2 f_{\text{R}}$ and an evolution time $t = 10/f_{\text{R}}$, are shown in Figure~\ref{fig:supp_filter_functions}. The peaks indicate the frequencies at which the CCD qubit is most sensitive to noise. For detuning noise in the CCD frame, the noise induces perturbations along both the longitudinal ($z$) axis and the transverse ($x$ and $y$) axes, causing dephasing and relaxation of the CCD qubit, respectively. The qubit is sensitive to detuning noise at frequencies $f_{\text{mod}}$ and $f_{\text{mod}} \pm \epsilon_{\text{mod}}$. Driving noise appears as transverse noise, leading to relaxation, with peak sensitivity at frequency $\epsilon_{\text{mod}}$. In contrast, modulation noise appears as longitudinal noise, resulting in dephasing, and the system is primarily sensitive to low-frequency components of modulation noise.

Using the filter function formalism, the coherence time of the CCD qubit can be understood as the weighted overlap between the filter function and the power spectral density of the noise; thus, the coherence enhancement of the CCD qubit depends on the power spectral density of each noise source as well as the modulation parameters $\epsilon_{\text{mod}}$ and $f_{\text{mod}}$. For phase-modulated microwaves, modulation noise is typically much smaller than the other noise sources and can be neglected. For quasi-static or $1/f^\alpha$ detuning noise and Rabi driving noise, increasing $f_{\text{mod}}$ (which requires increasing $f_{\text{R}}$) shifts the peak to higher frequencies, thereby reducing sensitivity to strong low-frequency noise and improving coherence. Increasing $\epsilon_{\text{mod}}$ reduces sensitivity to strong low-frequency driving noise; however, it simultaneously shifts the $f_{\text{mod}} - \epsilon_{\text{mod}}$ peak of the detuning noise filter function to lower frequencies, making the qubit more sensitive to low-frequency detuning noise. Therefore, careful tuning of the CCD parameters is necessary to optimize this trade-off.

\section{Numerical simulation of frequency detuning dependence}
\label{supp:numerical_simulation}

The ideal CCD theory assumes that the microwave drive is exactly resonant with the qubit transition and that the modulation frequency satisfies $f_{\text{mod}} = f_{\text{R}}$. Under these conditions, the expected dressed-qubit Rabi frequency is $\epsilon_{\text{mod}} = 0.2 f_{\text{R}} = 7.6\text{ MHz}$ for $f_{\text{R}} = 38\text{ MHz}$. Experimentally, however, both the dressed-qubit Rabi and Ramsey oscillation frequencies were measured to be approximately $8.8\text{ MHz}$, indicating a small deviation from the ideal resonant model.

\begin{figure*}[t]
  \centering
  \includegraphics[width=\textwidth]{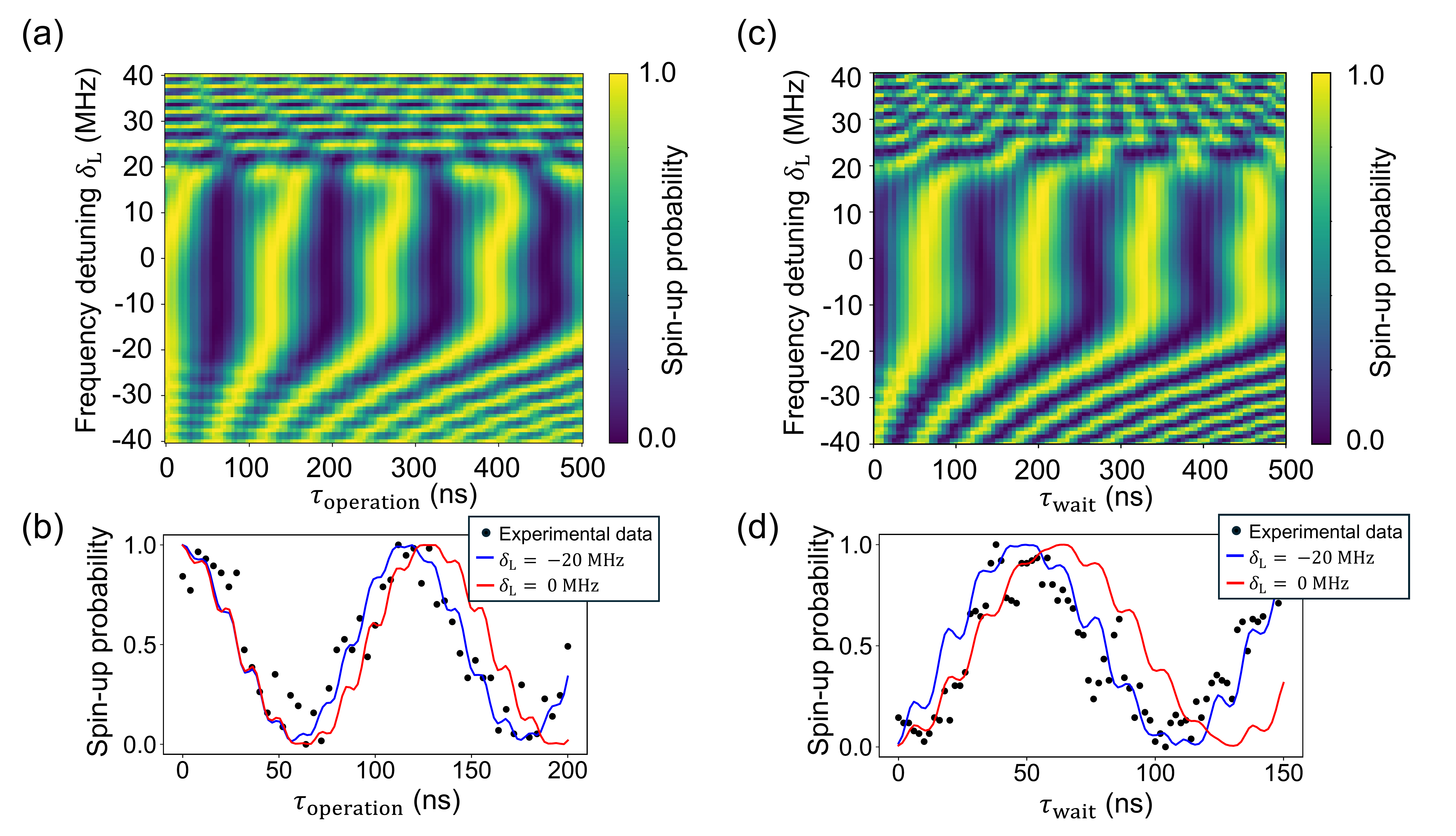}
  \caption{\textbf{Numerical simulation of the CCD-dressed qubit coherent oscillations as a function of driving frequency detuning $\delta_{\text{L}}$.} \textbf{(a)} and \textbf{(b)} show the Rabi oscillations while \textbf{(c)} and \textbf{(d)} show the Ramsey oscillations. The experimental data (black dots) are well reproduced by the numerical simulations using $\delta_{\text{L}} = -20\text{ MHz}$. The experimental current data were scaled and normalized for direct comparison with the simulated spin-up probability.}
  \label{fig:supp_detuning_simulation}
\end{figure*}

To investigate the origin of this discrepancy, we numerically solved the time-dependent Schrödinger equation in the lab frame while neglecting the longitudinal components of the effective field due to the SOC:
\begin{equation}
H_{\text{lab}} = \frac{h f_{\text{L}}}{2} \sigma_z + h f_{\text{R}} \cos\left[ 2\pi f_{\text{MW}} t + \phi_{\text{MW}} - \frac{2\epsilon_{\text{mod}}}{f_{\text{R}}} \sin(2\pi f_{\text{mod}} t - \theta_{\text{mod}}) \right] \sigma_x.
\label{eq:H_lab_detuning}
\end{equation}
All parameters were fixed to their experimentally determined values, with the microwave frequency $f_{\text{MW}}$ as the only adjustable parameter, corresponding to an effective qubit-drive detuning of $\delta_{\text{L}} = f_{\text{MW}} - f_{\text{L}}$. The numerical simulations show that an effective detuning of $\delta_{\text{L}} = -20\text{ MHz}$ quantitatively reproduces both the dressed-qubit Rabi and Ramsey oscillations, as shown in Figure~\ref{fig:supp_detuning_simulation}. The simulated oscillations exhibit weak wavy features because the modulation amplitude $\epsilon_{\text{mod}}$ is comparable to $f_{\text{mod}}$, resulting in a non-negligible counter-rotating term~\cite{kuno2026robust}.

The obtained effective detuning should be regarded as an effective parameter describing a deviation from the ideal resonant condition during the measurement rather than as a directly measured resonance-frequency shift. Such effective detuning may arise from slow variations of the qubit operating point caused by changes in the local electrostatic environment, including charge fluctuations, microwave-induced electrostatic drift, or microwave-driven rectification of the quantum-dot position due to the anharmonicity of the confinement potential. While the present simulations do not identify the microscopic origin of the detuning, they demonstrate that its inclusion is sufficient to quantitatively reproduce the experimentally observed dressed-qubit dynamics.

\end{document}